\documentstyle[12pt,axodraw]{article}

\catcode`@=12
\begin{document}
\thispagestyle{empty}
\begin{flushright} 
UCRHEP-T299\\ 
January 2001\
\end{flushright}
\vspace{0.5in}
\begin{center}
{\LARGE	\bf Canceling Quadratic Divergences in\\a Class of Two-Higgs-Doublet 
Models\\}
\vspace{1.5in}
{\bf Ernest Ma\\}
\vspace{0.2in}
{\sl Physics Department, University of California, Riverside, 
California 92521\\}
\vspace{1.5in}
\end{center}
\begin{abstract}\
The Newton-Wu conditions for the cancellation of quadratic divergences in 
a class of two-Higgs-doublet models are analyzed as to how they may be 
satisfied with a typical extension of the Standard Model of particle 
interactions.
\end{abstract}
\newpage
\baselineskip 24pt

In the Standard Model (SM) of particle interactions, there is one physical 
scalar Higgs boson $H$.  Its mass may be constrained by the requirement 
that the one quadratic divergence of the theory be canceled among its 
gauge, Yukawa, and quartic scalar couplings \cite{veltman}, i.e.
\begin{equation}
{3 \over 2} M_W^2 + {3 \over 4} M_Z^2 + {3 \over 4} m_H^2 = \sum_f N_f m_f^2,
\end{equation}
where the sum is over all SM fermions with $N_f$ its number of color 
degrees of freedom, i.e. $N_f = 3$ for quarks and $N_f = 1$ for leptons. 
The above has been written in terms of masses because all SM couplings 
are related to masses through the one vacuum expectation value (VEV) of the 
theory.  Given that $m_t = 174$ GeV and all other fermion masses are 
negligible compared to it, the prediction is then $m_H = 316$ GeV.

In the two-Higgs-doublet extension of the SM, there are 4 quadratic 
divergences and they are functions again of the gauge, Yukawa, and quartic 
scalar couplings.  They have been calculated previously by Newton and Wu 
\cite{newu}.  However, since there are now more possible quartic scalar 
couplings than masses, and the Yukawa terms are not uniquely determined 
(because of the different possible choices of which fermion couples to 
which scalar), the 4 conditions vary according to what specific 
two-Higgs-doublet extension is assumed.

In this note, the following Higgs potential for two doublets $\Phi_{1,2} = 
(\phi^+_{1,2}, \phi^0_{1,2})$ is considered:
\begin{eqnarray}
V &=& \mu_1^2 \Phi_1^\dagger \Phi_1 + \mu_2^2 \Phi_2^\dagger \Phi_2 + 
\mu_{12}^2 (\Phi_1^\dagger \Phi_2 + \Phi_2^\dagger \Phi_1) + {1 \over 2} 
\lambda_1 (\Phi_1^\dagger \Phi_1)^2 + {1 \over 2} \lambda_2 (\Phi_2^\dagger 
\Phi_2)^2 \nonumber \\ && + \lambda_3 (\Phi_1^\dagger 
\Phi_1)(\Phi_2^\dagger \Phi_2) + \lambda_4 (\Phi_1^\dagger \Phi_2)
(\Phi_2^\dagger \Phi_1) + {1 \over 2} \lambda_5 (\Phi_1^\dagger \Phi_2)^2 
+ {1 \over 2} \lambda_5^* (\Phi_2^\dagger \Phi_1)^2.
\end{eqnarray}
This is the most general form of $V$ under the assumption that a discrete 
odd-even symmetry ($\Phi_1$ is odd and $\Phi_2$ is even or vice versa) 
exists which is broken only by soft terms ($\mu_{12}^2$).  The utility 
of such a formulation in model building is widely recognized \cite{whepp2}.
The same odd-even symmetry restricts the Yukawa couplings of the SM fermions 
of each type to be associated with either $\Phi_1$ or $\Phi_2$, but not both. 
Under these assumptions, the 4 Newton-Wu conditions reduce to only two:
\begin{eqnarray}
{3 \over 2} M_W^2 + {3 \over 4} M_Z^2 + {v^2 \over 2} (3 \lambda_1 + 
2 \lambda_3 + \lambda_4) &=& {1 \over \cos^2 \beta} \sum_{f_1} N_{f_1} 
m_{f_1}^2, \\ {3 \over 2} M_W^2 + {3 \over 4} M_Z^2 + {v^2 \over 2} 
(3 \lambda_2 + 2 \lambda_3 + \lambda_4) &=& {1 \over \sin^2 \beta} \sum_{f_2} 
N_{f_2} m_{f_2}^2,
\end{eqnarray}
where $\langle \phi^0_{1,2} \rangle = v_{1,2}$, $v^2 = v_1^2 + v_2^2$, and 
$\tan \beta = v_2/v_1$.

There are 5 physical scalar particles.  Assuming $\lambda_5$ to be real for 
simplicity, their masses are given as follows \cite{dkmk}:
\begin{eqnarray}
m^2_{H^\pm} &=& -\mu_{12}^2 (\tan \beta + \cot \beta) - (\lambda_4 + 
\lambda_5) v^2, \\ m_A^2 &=& -\mu_{12}^2 (\tan \beta + \cot \beta) - 2 
\lambda_5 v^2,
\end{eqnarray}
with the other two ($m_{1,2}^2$) being the eigenvalues of the matrix
\begin{equation}
{\cal M}^2 = \left[ \begin{array} {c@{\quad}c} -\mu_{12}^2 \tan \beta + 2 
\lambda_1 v_1^2 & \mu_{12}^2 + 2(\lambda_3 + \lambda_4 + \lambda_5) v_1 v_2 \\ 
\mu_{12}^2 + 2(\lambda_3 + \lambda_4 + \lambda_5) v_1 v_2 & - \mu_{12}^2 
\cot \beta + 2 \lambda_2 v_2^2 \end{array} \right].
\end{equation}
In terms of $m_{1,2}^2$ and the mixing angle $\alpha$ which diagonalizes it, 
the above can be rewritten as
\begin{equation}
{\cal M}^2 = \left[ \begin{array} {c@{\quad}c} m_1^2 \cos^2 \alpha + m_2^2 
\sin^2 \alpha & (m_1^2 - m_2^2) \sin \alpha \cos \alpha \\ (m_1^2 - m_2^2) 
\sin \alpha \cos \alpha & m_2^2 \cos^2 \alpha + m_1^2 \sin^2 \alpha 
\end{array} \right].
\end{equation}
The conditions of Eqs.~(3) and (4) can now be formulated as
\begin{equation}
\left[ \begin{array} {c@{\quad}c} A_{11} & A_{12} \\ A_{21} & A_{22} 
\end{array} \right] \left[ \begin{array} {c} m_1^2 \\ m_2^2 \end{array} 
\right] = \left[ \begin{array} {c} C_1 \\ C_2 \end{array} \right],
\end{equation}
where
\begin{eqnarray}
A_{11} = {3 \cos^2 \alpha \over \cos^2 \beta} + {2 \sin \alpha \cos \alpha 
\over \sin \beta \cos \beta}, &~& A_{12} = {3 \sin^2 \alpha \over \cos^2 
\beta} - {2 \sin \alpha \cos \alpha \over \sin \beta \cos \beta}, \\ A_{21} = 
{3 \sin^2 \alpha \over \sin^2 \beta} + {2 \sin \alpha \cos \alpha \over 
\sin \beta \cos \beta}, &~& A_{22} = {3 \cos^2 \alpha \over \sin^2 \beta} 
- {2 \sin \alpha \cos \alpha \over \sin \beta \cos \beta},
\end{eqnarray}
and
\begin{eqnarray}
C_1 &=& {4 \over \cos^2 \beta} \sum_{f_1} N_{f_1} m_{f_1}^2 - 6 M_W^2 - 3 
M_Z^2 - 2 m_{H^\pm}^2 - m_A^2 - {\mu_{12}^2 \over \sin \beta \cos \beta} [1 + 
3 \tan^2 \beta], \\ C_2 &=& {4 \over \sin^2 \beta} \sum_{f_2} N_{f_2} 
m_{f_2}^2 - 6 M_W^2 - 3 M_Z^2 - 2 m_{H^\pm}^2 - m_A^2 - {\mu_{12}^2 \over 
\sin \beta \cos \beta} [1 + 3 \cot^2 \beta].
\end{eqnarray}

Let $\Phi_1$ couple to the $d, s, b$ quarks and $e, \mu, \tau$ 
leptons, and $\Phi_2$ to the $u, c, t$ quarks.  Then the main contribution to 
$C_1$ from fermions is $12 m_b^2/\cos^2 \beta$, and that to $C_2$ is 
$12 m_t^2/\sin^2 \beta$.  Hence $C_1 < 0$ and $C_2 > 0$ over most of the 
interesting parameter space.  Consider for example $m_{H^\pm} = 150$ GeV and 
$m_A = 100$ GeV.  Using
\begin{equation}
\left[ \begin{array} {c} m_1^2 \\m_2^2 \end{array} \right] = {1 \over A_{11} 
A_{22} - A_{21} A_{12}} \left[ \begin{array} {c@{\quad}c} A_{22} & -A_{12} \\ 
-A_{21} & A_{11} \end{array} \right] \left[ \begin{array} {c} C_1 \\ C_2 
\end{array} \right],
\end{equation}
$m_{1,2}$ may then be obtained as functions of $\mu_{12}^2$, $\tan \beta$, 
and $\tan \alpha$.  There are 2 special limits for $\mu_{12}^2$, namely 
zero and $\mu_{12}^2 = -m_A^2 \sin \beta \cos \beta$ for which $\lambda_5 
= 0$.  In the following numerical analysis, only these 2 limits will be 
considered.  Although the parameters $\tan \alpha$ and $\tan \beta$ are 
in principle independent, it turns out that solutions only exist if they 
are small and not too far apart in value.  Hence they will also be set 
equal here for simplicity.  In Table 1, $m_1$ and $m_2$ are shown for 
various values of $\tan \beta = \tan \alpha$ for the cases $\mu_{12}^2 = 0$ 
and $\lambda_5 = 0$.

\begin{center}
\begin{tabular} {|c|c|c|}
\hline 
 & $\mu_{12}^2 = 0$ & $\lambda_5 = 0$ \\
\hline 
$\tan \beta$ & $(m_1,m_2)$ [GeV] & $(m_1,m_2)$ [GeV] \\ 
\hline 
0.2 & (154,~355) & (172,~369) \\
0.3 & (150,~365) & (168,~378) \\
0.4 & (142,~380) & (162,~393) \\
0.5 & (129,~402) & (151,~414) \\
0.6 & (106,~435) & (131,~446) \\
0.7 & (39,~487) & (87,~497) \\
\hline 
\end{tabular}\\[10pt]
{Table 1. Values of $m_{1,2}$ from $\tan \beta = \tan \alpha$ for $\mu_{12}^2 
= 0$ and $\lambda_5 = 0$.}
\end{center}

It has thus been demonstrated that in a class of two-Higgs-doublet models 
defined by Eq.~(2), the cancellation of quadratic divergences is possible 
[i.e. Eqs.~(3) and (4)] with realistic values of the Higgs-boson masses, 
even though the right-hand-side of Eq.~(3) is negligible because of small 
Yukawa couplings. [Whereas $\lambda_1 > 0$ is required, $2 \lambda_3 + 
\lambda_4 < 0$ is allowed.]  In particular, the numerical values of Table 1 
show that $m_1$ of order 100 GeV \cite{higgs} is not impossible.

This work was supported in part by the U.~S.~Department of Energy
under Grant No.~DE-FG03-94ER40837.

\bibliographystyle{unsrt}

\end{document}